\documentclass[english,submission,copyright,creativecommons]{eptcs}
\usepackage[T1]{fontenc}
\usepackage[latin9]{inputenc}
\usepackage{array}
\usepackage{multirow}
\usepackage{graphicx}

\makeatletter

\providecommand{\tabularnewline}{\\}

\hypersetup{pdfusetitle,bookmarks=true,bookmarksnumbered=true,bookmarksopen=true,bookmarksopenlevel=2,breaklinks=false,pdfborder={0 0 0},backref=false,colorlinks=false}

\usepackage{etoolbox}

\newcommand\lyxeptcstitlerunning[2]{\title{#2}\ifstrempty{#1}{}{\def\titlerunning{#1}}}
\newcommand\lyxeptcsauthorrunning[2]{\author{#2}\ifstrempty{#1}{}{\def\authorrunning{#1}}}
\newcommand\lyxeptcsevent[1]{\providecommand{\event}{#1}}

\def\titlerunning{A Graphical ARA Model for O\&G Drilling Cybersecurity}
\def\authorrunning{A. Couce Vieira, S.H. Houmb \& D. Rios Insua}
\usepackage[strings]{underscore}

\makeatother

\usepackage{babel}
\begin{document}

\lyxeptcstitlerunning{}{A Graphical Adversarial Risk Analysis Model for Oil and Gas Drilling
Cybersecurity}

\lyxeptcsauthorrunning{}{Aitor Couce Vieira \qquad{}\qquad{} Siv Hilde Houmb \institute{Secure-NOK AS\\
Stavanger, Norway}\email{\quad{}aitorcouce@securenok.com\quad{}\qquad{}sivhoumb@securenok.com}\and David
Rios Insua \institute{Royal Academy of Sciences\\
Madrid, Spain}\email{david.rios@urjc.es} }

\lyxeptcsevent{GraMSec 2014}
\maketitle
\begin{abstract}
Oil and gas drilling is based, increasingly, on operational technology,
whose cybersecurity is complicated by several challenges. We propose
a graphical model for cybersecurity risk assessment based on Adversarial
Risk Analysis to face those challenges. We also provide an example
of the model in the context of an offshore drilling rig. The proposed
model provides a more formal and comprehensive analysis of risks,
still using the standard business language based on decisions, risks,
and value.
\end{abstract}

\section{Introduction}

Operational technology (OT) refers to \textquotedblleft{}hardware
and software that detects or causes a change through the direct monitoring
and/or control of physical devices, processes and events in the enterprise\textquotedblright{}
\cite{gartner}. It includes technologies such as SCADA systems. Implementing
OT and information technology (IT) typically lead to considerable
improvements in industrial and business activities, through facilitating
the mechanization, automation, and relocation of activities in remote
control centers. These changes usually improve the safety of personnel,
and both the cost-efficiency and overall effectiveness of operations.

The oil and gas industry (O\&G) is increasingly adopting OT solutions,
in particular offshore drilling, through drilling control systems
(drilling CS) and automation, which have been key innovations over
the last few years. The potential of OT is particularly relevant for
these activities: centralizing decision-making and supervisory activities
at safer places with more and better information; substituting manual
mechanical activities by automation; improving data through better
and near real-time sensors; and optimizing drilling processes. In
turn, they will reduce rig crew and dangerous operations, and improve
efficiency in operations, reducing operating costs (typically of about
\$300,000 per day).

Since many of the involved OT employed in O\&G are currently computerized,
they have become a major potential target for cyber attacks \cite{Shauk2013c},
given their economical relevance, with large stakes at play. Indeed,
we may face the actual loss of large oil reserves because of delayed
maneuvers, the death of platform personnel, or potential large spills
with major environmental impact with potentially catastrophic consequences.
Moreover, it is expected that security attacks will soon target several
production installations simultaneously with the purpose of sabotaging
production, possibly taking advantage of extreme weather events, and
attacks oriented towards manipulating or obtaining data or information.
Cybersecurity poses several challenges, which are enhanced in the
context of operational technology. Such challenges are sketched in
the following section.

\subsection{Cybersecurity Challenges in Operational Technology}

Technical vulnerabilities in operational technology encompass most
of those related with IT vulnerabilities \cite{byres2004myths}, complex
software \cite{DoDDefSci2013}, and integration with external networks
\cite{giani2009viking}. There are also and specific OT vulnerabilities
\cite{zhu2011taxonomy,brenner2013eyes}. However, OT has also strengths
in comparison with typical IT systems employing simpler network dynamics.

Sound organizational cybersecurity is even more important with OT
given the risks that these systems bring in. Uncertainties are considerable
in both economical and technical sense \cite{anderson2010security}.
Therefore better data about intrusion attempts are required for improving
cybersecurity \cite{pfleeger2008cybersecurity}, although gathering
them is difficult since organizations are reluctant about disclosing
such information \cite{ten2008vulnerability}.

More formal approaches to controls and measures are needed to deal
with advanced threat agents such as assessing their attack patterns
and behavior \cite{hutchins2011intelligence} or implementing intelligent
sensor and control algorithms \cite{cardenas2008research}. An additional
problem is that metrics used by technical cybersecurity to evaluate
risks usually tell little to those evaluating or making-decisions
at the organizational cybersecurity level. Understanding the consequences
of a cyber attack to an OT system is difficult. They could lead to
production losses or the inability to control a plant, multimillion
financial losses, and even impact stock prices \cite{byres2004myths}.
One of the key problems for understanding such consequences is that
OT systems are also cyber-physical systems (CPS) encompassing both
computational and complex physical elements \cite{thomas2013bad}.

Risk management is also difficult in this context \cite{mulligan2011doctrine}.
Even risk standards differ on how to interpret risk: some of them
assess the probabilities of risk, others focus on the vulnerability
component \cite{hutchins2011intelligence}. Standards also tend to
present oversimplifications that might alter the optimal decision
or a proper understanding of the problem, such as the well-known shortcomings
of the widely employed risk matrices \cite{cox2008matrix}.

Cyber attacks are the continuation of physical attacks by digital
means. They are less risky, cheaper, easier to replicate and coordinate,
unconstrained by distance \cite{cardenas2009challenges}, and they
could be oriented towards causing high impact consequences \cite{DoDDefSci2013}.
It is also difficult to measure data related with attacks such as
their rate and severity, or the cost of recovery \cite{anderson2010security}.
Examples include Stuxnet \cite{brenner2013eyes}, Shamoon \cite{brenner2013eyes},
and others \cite{cardenas2008research}. Non targeted attacks could
be a problem also.

Several kinds of highly skilled menaces of different nature (e.g.,
military, hacktivists, criminal organizations, insiders or even malware
agents) can be found in the cyber environment \cite{DoDDefSci2013},
all of them motivated and aware of the possibilities offered by OT
\cite{byres2004myths}. Indeed, the concept Advanced Persistent Threat
(APT) has arisen to name some of the threats \cite{Ltd2011}. The
diversity of menaces could be classified according their attitude,
skill and time constraints \cite{dantu2007classification}, or by
their ability to exploit, discover or even create vulnerabilities
on the system \cite{DoDDefSci2013}. Consequently, a sound way to
face them is profiling \cite{atzeni2011here} and treating \cite{li2009botnet}
them as adversarial actors.

\subsection{Related Work Addressing the Complexities of Cybersecurity Challenges}

Several approaches have been proposed to model attackers and attacks,
including stochastic modelling \cite{muehrcke2010behavior,sallhammar2007stochastic},
attack graph models \cite{kotenko2006attack} and attack trees \cite{mauw2006foundations},
models of directed and intelligent attacks \cite{ten2008vulnerability};
models based on the kill chain attack phases \cite{hutchins2011intelligence},
models of APT attack phases \cite{Ltd2011}, or even frameworks incorporating
some aspects of intentionality or a more comprehensive approach to
risk such as CORAS \cite{lund2011model} or ADVISE \cite{Advise2013}.

Game theory has provided insights concerning the behavior of several
types of attackers \textendash{} such as cyber criminal APTs \textendash{}
and how to deal with them. The concept of incentives can unify a large
variety of agent intents, whereas the concept of utility can integrate
incentives and costs in such a way that the agent objectives can be
modeled in practice \cite{liu2005incentive}. Important insights from
game theory are that the defender with lowest protection level tends
to be a target for rational attackers \cite{Johnson2011}, that defenders
tend to under-invest in cybersecurity \cite{amin2011interdependence},
and that the attacker\textquoteright{}s target selection is costly
and hard, and thus it needs to be carefully carried on \cite{florencio2013all}.
In addition to such general findings, some game-theoretic models exist
for cybersecurity or are applicable to it, modelling static and dynamic
games in all information contexts \cite{roy2010survey}. However,
game-theoretic models have their limitations \cite{hamilton2002challenges,roy2010survey}
such as limited data, the difficulty to identify the end goal of the
attacker, the existence of a dynamic and continuous context, and that
they are not scalable to the complexity of real cybersecurity problems
in consideration. Moreover, from the conceptual point of view, they
require common knowledge assumptions that are not tenable in this
type of applications.

Additionally, several Bayesian models have been proposed for cybersecurity
risk management such as a model for network security risk analysis
\cite{xie2010using}; a model representing nodes as events and arcs
as successful attacks \cite{dantu2007classification}; a dynamic Bayesian
model for continuously measuring network security \cite{frigault2008measuring};
a model for Security Risk Management incorporating attacker capabilities
and behavior \cite{dantu2009network}: or models for intrusion detection
systems (IDS) \cite{balchanos2012probabilistic}. However, these models
require forecasting attack behavior which is hard to come by.

Adversarial Risk Analysis (ARA) \cite{rios2009adversarial} combine
ideas from Risk Analysis, Decision Analysis, Game-Theory, and Bayesian
Networks to help characterizing the motivations and decisions of the
attackers. ARA is emerging as a main methodological development in
this area \cite{merrick2011comparative}, providing a powerful framework
to model risk analysis situations with adversaries ready to increase
our threats. Applications in physical security may be seen in \cite{sevillano2012adversarial}.

\subsection{Our Proposal}

The challenges that face OT, cybersecurity and the O\&G sector create
a need of a practical, yet rigorous approach, to deal with them. Work
related with such challenges provides interesting insights and tools
for specific issues. However, more formal but understandable tools
are needed to deal with such problems from a general point of view,
without oversimplifying the complexity underlying the problem. We
propose a model for cybersecurity risk decisions based on ARA, taking
into account the attacker behavior. Additionally, an application of
the model in drilling cybersecurity is presented, tailored to decision
problems that may arise in offshore rigs employing drilling CS.

\section{Model}

\subsection{Introduction to Adversarial Risk Analysis}

ARA aims at providing one-sided prescriptive support to one of the
intervening agents, the Defender (she), based on a subjective expected
utility model, treating the decisions of the Attacker (he) as uncertainties.
In order to predict the Attacker\textquoteright{}s actions, the Defender
models her decision problem and tries to assess her probabilities
and utilities but also those of the Attacker, assuming that the adversary
is an expected utility maximizer. Since she typically has uncertainty
about those, she models it through random probabilities and uncertainties.
She propagates such uncertainty to obtain the Attacker's optimal random
attack, which she then uses to find her optimal defense.

ARA enriches risk analysis in several ways. While traditional approaches
provide information about risk to decision-making, ARA integrates
decision-making within risk analysis. ARA assess intentionality thoroughly,
enabling the anticipation and even the manipulation of the Attacker
decisions. ARA incorporates stronger statistical and mathematical
tools to risk analysis that permit a more formal approach of other
elements involved in the risk analysis. It improves utility treatment
and evaluation. Finally, an ARA graphical model improves the understandability
of complex cases, through visualizing the causal relations between
nodes.

The main structuring and graphical tool for decision problems are
Multi-Agent Influence Diagrams (MAID), a generalization of Bayesian
networks. ARA is a decision methodology derived from Influence Diagrams,
and it could be structured with the following basic elements:
\begin{itemize}
\item \emph{Decisions or Actions}. Set of alternatives which can be implemented
by the decision makers. They represent what one can do. They are characterized
as decision nodes (rectangles).
\item \emph{Uncertain States}. Set of uncontrollable scenarios. They represent
what could happen. They are characterized as uncertainty nodes (ovals).
\item \emph{Utility and Value}. Set of preferences over the consequences.
They represent how the previous elements would affect the agents.
They are characterized as value nodes (rhombi).
\item \emph{Agents}. Set of people involved in the decision problem: decision
makers, experts and affected people. In this context, there are several
agents with opposed interests. They are represented through different
colors.
\end{itemize}
We describe now the basic MAID that may serve as a template for cybersecurity
problems in O\&G drilling CS, developed using GeNIe \cite{genie}.

\subsection{Graphical Model}

Our model captures the Defender cybersecurity main decisions prior
to an attack perpetrated by an APT, which is strongly \textquotedblleft{}business-oriented\textquotedblright{}.
Such cyber criminal organization behavior suits utility-maximizing
analysis, as it pursues monetary gains. A sabotage could also be performed
by this type of agents, and they could be hired to make the dirty
job for a foreign power or rival company. We make several assumptions
in the Model, to make it more synthetic:
\begin{itemize}
\item We assume one Defender. The Attacker\textquoteright{}s nodes do not
represent a specific attacker, but a generalization of potential criminal
organizations that represent business-oriented APTs, guided mostly
by monetary incentives.
\item We assume an atomic attack (the attacker makes one action), with several
consequences, as well as several residual consequences once the risk
treatment strategy is selected.
\item The Defender and Attacker costs are deterministic nodes.
\item We avoid detection-related activities or uncertainties to simplify
the Model. Thus, the attack is always detected and the Defender is
always able to respond to it.
\item The scope of the Model is an assessment activity prior to any attack,
as a risk assessment exercise to support incident handling planning.
\item The agents are expected utility maximizers.
\item The Model is discrete.
\end{itemize}
By adapting the proposed template in Figure 1, we may generalize most
of the above assumptions to the cases at hand.

\clearpage

\begin{center}

\includegraphics{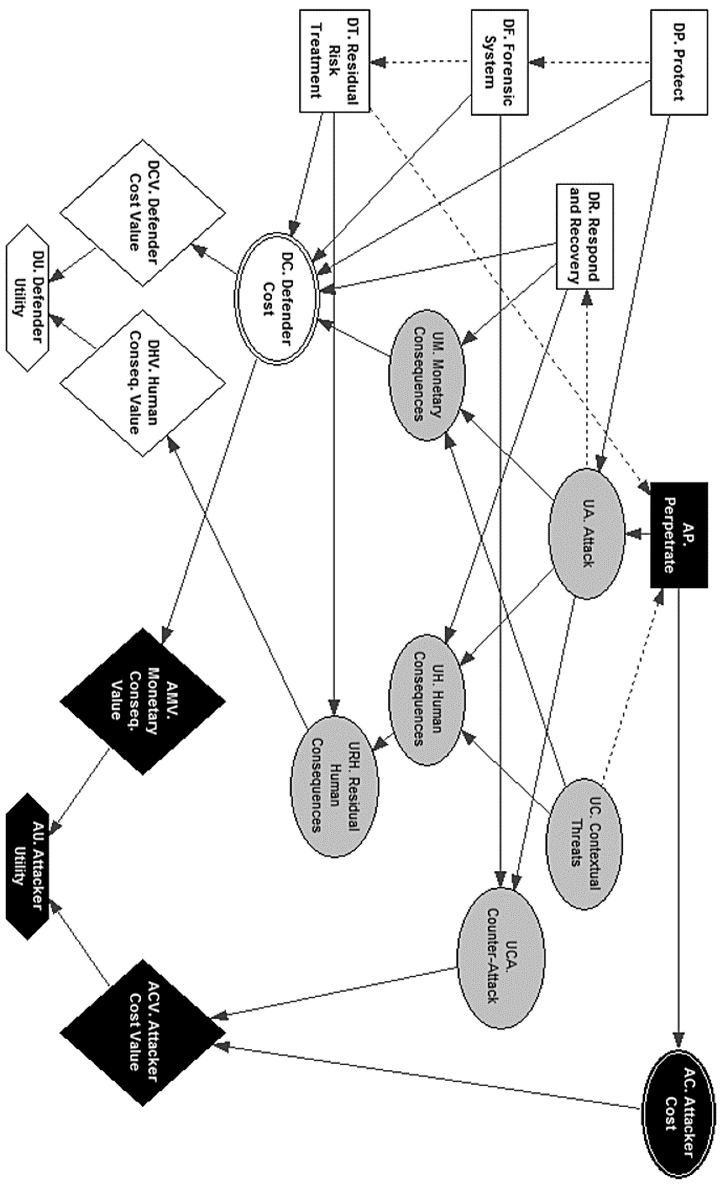}

\textbf{Figure 1}. MAID of the ARA Model for O\&G drilling cybersecurity.

\end{center}

\subsubsection{Defender Decision and Utility Nodes}

The Defender nodes, in white, are:
\begin{itemize}
\item \emph{Protect} (\emph{DP}) decision node. The Defender selects among
security measures portfolios to increase protection against an Attack,
e.g., access control, encryption, secure design, firewalls, or personal
training and awareness.
\item \emph{Forensic System} (\emph{DF}) decision node. The Defender selects
among different security measures portfolios that may harm the Attacker,
e.g., forensic activities that enable prosecution of the Attacker.
\item \emph{Residual Risk Treatment} (\emph{DT}) decision node. This node
models Defender actions after the assessment of other decisions made
by the Defender and the Attacker. They are based on the main risk
treatment strategies excluding risk mitigation, as they are carried
on through the Protect and the Respond and Recovery nodes: avoiding,
sharing, or accepting risk. This node must be preceded by the Protect
defender decision node, and it must precede the Attack uncertainty
node (the residual risk assessment is made in advance).
\item \emph{Respond and Recovery} (\emph{DR}) decision node. The Defender
selects between different response and recovery actions after the
materialization of the attack, trying to mitigate the attack consequences.
This will depend on the attack uncertainty node.
\item \emph{Defender Cost} (\emph{DC}) deterministic node. The costs of
the decisions made by the Defender are deterministic, as well as the
monetary consequences of the attack (the uncertainty about such consequences
is solved in the Monetary Consequences node). In a more sophisticated
model, most of the costs could be modeled as uncertain nodes. This
node depends on all decision nodes of the Defender and the Monetary
Consequences uncertainty node.
\item \emph{Value Nodes} (\emph{DCV} and \emph{DHV}). The Defender evaluates
the consequences and costs, taking into account her risk attitude.
They depend on the particular nodes evaluated at each Value node.
\item \emph{Utility Nodes} (\emph{DU}). This node merges the Value nodes
of the Defender. It depends on the Defender\textquoteright{}s Value
nodes.
\end{itemize}
The Decision nodes are adapted to the typical risk management steps,
incorporating ways of evaluating managing sound organizational cybersecurity
strategy, which takes into account the business implications of security
controls, and prepare the evaluation of risk consequences. Related
work (Section 1.2) on security costs and investments could incorporate
further complexities underlying the above nodes.

\subsubsection{Attacker Decision and Utility Nodes}

The Attacker nodes, in black, are:
\begin{itemize}
\item \emph{Perpetrate} (AP) decision node. The {[}generic{]} Attacker decides
whether he attacks or not. It could be useful to have a set of options
for a same type of attack (e.g., preparing a quick and cheap attack,
or a more elaborated one with higher probabilities of success). It
should be preceded by the Protect and Residual Risk Treatment decision
nodes, and might be preceded by the Contextual Threat node (in case
the Attacker observes it).
\item \emph{Attacker Cost} (\emph{AC}) deterministic node. Cost of the Attacker
decisions. Preceded by the Perpetrate decision node.
\item \emph{Value Nodes} (\emph{AMV} and \emph{ACV}). The Attacker evaluates
the different consequences and costs, taking into account his risk
attitude. They depend on the deterministic or uncertainty nodes evaluated
at each Value node.
\item \emph{Utility Nodes} (\emph{AU}). It merges the Value nodes of the
Attacker to a final set of values. It must depend on the Attacker\textquoteright{}s
Value nodes.
\end{itemize}
These nodes help in characterizing the Attacker, avoiding the oversimplification
of other approaches. Additionally, the Defender has uncertainty about
the Attacker probabilities and utilities. This is propagated over
their nodes, affecting the Attacker expected utility and optimal alternatives,
which are random. Such distribution over optimal alternatives is our
forecast for the Attacker's actions.

\subsubsection{Uncertainty Nodes}

The uncertainty nodes in grey are:
\begin{itemize}
\item \emph{Contextual Threats} (\emph{UC}) uncertainty node. Those threats
(materialized or not) present during the Attack. The Attacker may
carry out a selected opportunistic Attack (e.g. hurricanes or a critical
moment during drilling).
\item \emph{Attack} (\emph{UA}) uncertainty node. It represents the likelihood
of the attack event, given its conditioning nodes. It depends on the
Perpetrate decision node, and on the Protect decision node.
\item \emph{Consequences} (\emph{UM} and \emph{UV}) uncertainty node. It
represents the likelihood of different consequence levels that a successful
attack may lead to. They depend on the Attack and Contextual Threat
uncertainty nodes, and on the Respond and Recovery decision node.
\item \emph{Residual Consequences} (\emph{URH}) uncertainty node. It represents
the likelihood of different consequence levels after applying residual
risk treatment actions. They depend on the Consequence node modelling
the same type of impact (e.g., human, environmental, or reputation).
\item \emph{Counter-Attack} (\emph{UCA}) uncertainty node. Possibility,
enabled by a forensic system, to counter-attack and cause harm to
the Attacker. Most of the impacts may be monetized. It depends on
the Forensic System decision node.
\end{itemize}
Dealing with the uncertainties and complexities and obtaining a probability
distribution for these nodes could be hard. Some of the methodologies
and findings proposed in the sections 1.1 and 1.2 are tailored to
deal with some of these complexities. Using them, the Model proposed
in this paper could lead to limit the uncertainties in cybersecurity
elements such as vulnerabilities, controls, consequences, attacks,
attacker behavior, and risks. This will enable achieving simplification,
through the proposed Model, without limiting the understanding of
the complexities involved, and a sounder organizational cybersecurity.

\section{Example}

We present a numerical example of the previous Model tailored to a
generic decision problem prototypical of a cybersecurity case that
may arise in O\&G offshore rig using drilling CS. The model specifies
a case in which the driller makes decisions to prevent and respond
to a cyber attack perpetrated by a criminal organization with APT
capabilities, in the context of offshore drilling and drilling CS.
The data employed in this example are just plausible figures helpful
to provide an overview of the problems that drilling cybersecurity
faces. Carrying on the assessment that the Model enables may be helpful
for feeding a threat knowledge base, incident management procedures
or incident detection systems.

The context is that of an offshore drilling rig, a floating platform
with equipment to drill a well through the seafloor, trying to achieve
a hydrocarbon reservoir. Drilling operations are dangerous and several
incidents may happen in the few months (usually between 2 or 4) that
the entire operation may last. As OT, drilling CS may face most of
the challenges presented in Section 1.1 (including being connected
to Enterprise networks, an entry path for attackers) in the context
of high-risk incidents that occur in offshore drilling.

\subsection{Agent Decisions}

\subsubsection{Defender Decisions}

The Defender has to make three decisions in advance of the potential
attack. In the Protect decision node (DP), the Defender must decide
whether she invests in additional protection: if the Defender implements
additional protective measures, the system will be less vulnerable
to attacks. In the Forensic System decision node (DF), the Defender
must decide whether she implements a forensic system or not. Implementing
it enables the option of identifying the Attacker and pursuing legal
or counter-hacking actions against him. The Residual Risk Treatment
decision node (DT) represents additional risk treatment strategies
that the Defender is able to implement: avoiding (aborting the entire
drilling operation to elude the attack), sharing (buying insurance
to cover the monetary losses of the attack), and accepting the risk
(inheriting all the consequences of the attack, conditional on to
the mitigation decisions of DP, FD, and DR).

Additionally, the Respond and Recovery decision node (DR) represents
the Defender\textquoteright{}s decision between continuing and stopping
the drilling operations as a reaction to the attack. Continuing the
drilling may lead to worsen the consequences of the attack, whereas
stopping the drilling will incur in higher costs due to holding operations.
This is a major issue for drilling CS. In general, critical equipment
should not be stopped, since core operations or even the safety of
the equipment or the crew may be compromised.

\subsubsection{Attacker Decisions}

For simplicity, in the Perpetrate decision node (AP) the Attacker
decides whether he perpetrates the attack or not, although further
attack options could be added. In this example, the attack aims at
manipulating the devices directly under control of physical systems
with the purpose of compromising drilling operations or harming equipment,
the well, the reservoir, or even people.

\subsection{Threat Outcomes and Uncertainty}

\subsubsection{Outcomes and Uncertainty during the Incident}

The Contextual Threats uncertainty node (UC) represents the existence
of riskier conditions in the drilling operations (e.g., bad weather
or one of the usual incidents during drilling), which can clearly
worsen the consequences of the attack. In this scenario, the Attacker
is able to know, to some extent, these contextual threats (e.g., a
weather forecast, a previous hacking in the drilling CS that permits
the attacker to read what is going on in the rig).

The Attack uncertainty node (UA) represents the chances of the Attacker
of causing the incident. If the Attacker decides not to execute his
action, no attack event will happen. However, in case of perpetration,
the chances of a successful attack will be lower if the Defender invests
in protective measures (DC node). An additional uncertainty arises
in case of materialization of the attack: the possibility to identify
and counter-attack the node, represented by the Counter-Attack uncertainty
node (UCA).

If the attack happens, the Defender will have to deal with different
consequence scenarios. The Monetary (UM) and Human Consequences (UH)
nodes represent the chances of different consequences or impact levels
that the Defender may face. The monetary consequences refer to all
impacts that can be measured as monetary losses, whereas human consequences
represent casualties that may occur during an incident or normal operations.
However, the Defender has the option to react to the attack by deciding
whether she continues or stops the drilling (DR node). If the Defender
decides to stop, there will be lower chances of casualties and lower
chances of worst monetary consequences (e.g., loss of assets or compensations
for injuries or deaths), but she will have to assume the costs of
keeping the rig held (one day in our example) to deal with the cyber
threat.

\subsubsection{Outcomes and Uncertainty in Risk Management Process}

The previous uncertainties appear after the Attacker's decision to
attack or not. The Defender faces additional relevant uncertainties.
She must make a decision between avoiding, sharing, or accepting the
risk (DT node). Such decision will determine the final or residual
consequences. The final monetary consequences are modeled through
the Defender Cost deterministic node (DC node), whose outcome represents
the cost of different Defender decisions (nodes DP, DF, DT, and DR).
In case of accepting or sharing the risk, the outcome of the DC node
will also inherit the monetary consequences of the attack (UM node).
Similarly, the outcome of the Residual Human Consequences uncertainty
node (URH) is conditioned by the risk treatment decisions (DC node)
and, in case of accepting or sharing the risk, it will inherit the
human consequences of the attack (UH node). If the Defender decides
to avoid the risk, she will assume the cost of avoiding the entire
drilling operations and will cause that the crew face a regular death
risk rather than the higher death risk of offshore operations. If
the Defender shares the risk, she will assume the same casualties
as in UH and a fixed insurance payment, but she will avoid paying
high monetary consequences. Finally, in case the Defender accepts
the risk, she will inherit the consequences from the UM and UH nodes.

The Attacker Cost deterministic node (AC) provides the costs (non-uncertain
by assumption) of the decision made by the Attacker. Since he only
has two decisions (perpetrate or not), the node has only two outcomes:
cost or not. This node could be eliminated, but we keep it to preserve
the business semantics within the graphical model.

\subsection{Agent Preferences}

The Defender aims at maximizing her expected utility, with the utility
function being additive, through the Defender Utility node (DU). The
Defender key objective is minimizing casualties, but he also considers
minimizing his costs (in this example we assume she is risk-neutral).
Each objective has its own weight in the utility function.

The objective of the Attacker is to maximize his expected utility,
represented by an additive utility function, through the Attacker
Utility node (AU). The Attacker key objective is maximizing the monetary
consequences for the Defender. We assume that he is risk-averse towards
this monetary impact (he prefers ensuring a lower impact than risking
the operations trying to get a higher impact). He also considers minimizing
his costs (i.e., being identified and perpetrating the attack). Each
of these objectives has its own weight in the utility function, and
its own value function. The Attacker does not care about eventual
victims.

\subsection{Uncertainty about the Opponent Decisions}

The Attacker is able to know to some extent the protective decisions
of the defender (DP node), gathering information while he tries to
gain access to the drilling CS. While knowing if the Defender avoided
the risk (avoiding all the drilling operations) is easy, knowing if
the Defender chose between sharing or accepting the risk is difficult.
The most important factor, the decision between continue or stop drilling
in case of an attack, could be assessed by observing the industry
or company practices. The Defender may be able to assess also how
frequent similar attacks are, or how attractive the drilling rig is
for this kind of attacker. In ARA, and from the Defender perspective,
the AP node would be an uncertainty node whose values should be provided
by assessing the probabilities of the different attack actions, through
analyzing the decision problem from the Attacker perspective and obtaining
his random optimal alternative.

\subsection{Example Values}

An annex provides the probability tables of the different uncertainty
nodes employed to simulate the example in Genie (Tables 1 to 7). It
also provides the different parameters employed in the utility and
value functions (Tables 8 to 10). Additionally, the \textquotedblleft{}risk-averse\textquotedblright{}
values for AMV are obtained with $AMV=\sqrt[3]{\frac{DC}{10^{7}}}$;
the \textquotedblleft{}risk-neutral\textquotedblright{} values for
DCV are obtained with $DCV=1-\frac{DC}{10^{7}}$; and, the values
for DHV are 0 in case of victims and 1 in case of no victims.

\subsection{Evaluation of Decisions}

Based on the solution of the example, we may say that the Attacker
should not perpetrate his action in case he believes the Defender
will avoid or share the risk. However, the Attacker may be interested
in perpetrating his action in case he believes that the Defender is
accepting the risk. Additionally, the less preventive measures the
Defender implements (DP and DT nodes), the more motivated the Attacker
would be (if he thinks the Defender is sharing the risk). The Attacker\textquoteright{}s
expected utility is listed in Table 11 in the Annex. The Defender
will choose in this example not to implement additional protection
(DP node) without a forensic system (DF node). If the Defender believes
that she is going to be attacked, then she would prefer sharing the
risk (DT node) and stop drilling after the incident (DR node). In
case she believes that there will be no attack, she should accept
the risk and continue drilling. The Defender\textquoteright{}s expected
utilities are listed in Table T12 in Annex.

Thus, the Defender optimal decisions create a situation in which the
Attacker is more interested in perpetrating the attack. Therefore,
to affect the Attacker\textquoteright{}s behavior, the Defender should
provide the image that her organization is concerned with safety,
and especially that it is going to share risks. On the other hand,
if the Attacker perceives that the Defender pays no attention to safety
or that she is going to accept the risk, he will try to carry on his
attack. The ARA solution for the Defender is the following:
\begin{enumerate}
\item Assess the problem from the point of view of the Attacker. The DT
and DR nodes are uncertainty nodes since that Defender decisions are
uncertain for the Attacker. The Defender must model such nodes in
the way that she thinks the Attacker models such uncertainties. In
general, perpetrating an attack is more attractive in case the Attacker
strongly believes that the Defender is going to accept the risk or
is going to continue drilling.
\item Once forecasted the Attacker\textquoteright{}s decision, the Defender
should choose between sharing and accepting the risk. Accepting the
risk in case of no attack is better than sharing the risk, but accepting
the risk in case of attack is worse.
\end{enumerate}
Thus, the key factor for optimizing the decision of the Defender are
her estimations on the uncertainty nodes that represent the DT and
DR nodes for the attacker. Such nodes will determine the Attacker
best decision, and this decision the Defender best decision.

\section{Conclusions and Further Work}

We have presented the real problem and extreme consequences that OT
cybersecurity in general, and drilling cybersecurity in particular,
are facing. We also explained some of the questions that complicate
cybersecurity, especially in OT systems. The proposed graphical model
provides a more comprehensive, formal and rigorous risk analysis for
cybersecurity. It is also a suitable tool, able of being fed by, or
compatible with, other more specific models such as those explained
in Section 1.

Multi-Agent Influence Diagrams provide a formal and understandable
way of dealing with complex interactive issues. In particular, they
have a high value as business tools, since its nodes translate the
problem directly into business language: decisions, risks, and value.
Typical tools employed in widely used risk standards, such as risk
matrices, oversimplify the problem and limit understanding. The proposed
ARA-based model provides a business-friendly interpretation of a risk
management process without oversimplifying its underlying complexity.

The ARA approach permits us to include some of the findings of game
theory applied to cybersecurity, and it also permits to achieve new
findings. The model provides an easier way to understand the problem
but it is still formal since the causes and consequences in the model
are clearly presented, while avoiding common knowledge assumptions
in game theory.

Our model presents a richer approach for assessing risk than risk
matrices, but it still has the security and risk management language.
In addition, it is more interactive and modular, nodes can be split
into more specific ones. The proposed model can still seem quite formal
to business users. However, data can be characterized using ordinal
values (e.g., if we only know that one thing is more likely/valuable
than other), using methods taken from traditional risk management,
employing expert opinion, or using worst case figures considered realistic.
The analysis would be poorer but much more operational.

Using the nodes of the proposed model as building blocks, the model
could gain in comprehensiveness through adding more attackers or attacks,
more specific decision nodes, more uncertainty nodes, or additional
consequence nodes, such as environmental impact or reputation. Other
operations with significant business interpretation can be done, such
as sensitivity analysis (how much the decision-makers should trust
a figure) or strength of the influence analysis (which are the key
elements).

Its applicability is not exempt of difficulties and uncertainties,
but in the same way than other approaches. Further work is needed
to verify and validate the model and its procedures (in a similar
way to the validation of other ARA-based models\cite{RiosInsua2013}),
and to identify the applicability and usability issues that may arise.
The model could gain usability through mapping only the relevant information
to decision-makers (roughly, decisions and consequences) rather than
the entire model.

\let\thefootnote\relax\footnote{\textbf{Acknowledgments}
\\- Work supported by the EU's FP7 Seconomics project 285223
\\- David Rios Insua grateful to the support of the MINECO, Riesgos project and the Riesgos-CM program
}

\clearpage

\bibliographystyle{eptcs}
\bibliography{ARAOGgramsec}

\appendix

\section*{Appendix: Tables with Example Data}

\begin{center}

\renewcommand{\arraystretch}{0.75}

\textbf{\scriptsize{\\Table T1}}{\scriptsize{. Probability table
for UC node.}}{\scriptsize \par}

{\scriptsize{}}%
\begin{tabular}{|c|c|}
\hline 
\emph{\scriptsize{Riskier conditions}} & {\scriptsize{30\%}}\tabularnewline
\hline 
\emph{\scriptsize{Normal conditions}} & {\scriptsize{70\%}}\tabularnewline
\hline 
\end{tabular}{\scriptsize \par}

\textbf{\scriptsize{\\Table T2}}{\scriptsize{. Probability table
for UA node.}}{\scriptsize \par}

{\scriptsize{}}%
\begin{tabular}{|c|c|c|c|c|}
\hline 
\emph{\scriptsize{Attacker's Perpetrate decision}} & \multicolumn{2}{c|}{\emph{\scriptsize{Perpetrate}}} & \multicolumn{2}{c|}{\emph{\scriptsize{No perpetrate}}}\tabularnewline
\hline 
\emph{\scriptsize{Defender's Protect decision}} & \emph{\scriptsize{Additional protection}} & \emph{\scriptsize{Non additional protection}} & \emph{\scriptsize{Additional protection}} & \emph{\scriptsize{Non additional protection}}\tabularnewline
\hline 
\hline 
\emph{\scriptsize{Attack event}} & {\scriptsize{5\%}} & {\scriptsize{40\%}} & {\scriptsize{0\%}} & {\scriptsize{0\%}}\tabularnewline
\hline 
\emph{\scriptsize{No attack event}} & {\scriptsize{95\%}} & {\scriptsize{60\%}} & {\scriptsize{100\%}} & {\scriptsize{100\%}}\tabularnewline
\hline 
\end{tabular}{\scriptsize \par}

\textbf{\scriptsize{\\Table T3.}}{\scriptsize{ Probability table
for UM node.}}{\scriptsize \par}

{\scriptsize{}}%
\begin{tabular}{|>{\centering}p{3.5cm}|>{\centering}m{1.1cm}|>{\centering}m{1.1cm}|>{\centering}m{1.1cm}|>{\centering}m{1.1cm}|>{\centering}m{1.1cm}|>{\centering}m{1.1cm}|>{\centering}m{1.1cm}|>{\centering}m{1.1cm}|}
\hline 
\emph{\scriptsize{Attack event}} & \multicolumn{4}{c|}{\emph{\scriptsize{Attack}}} & \multicolumn{4}{c|}{\emph{\scriptsize{No attack}}}\tabularnewline
\hline 
\emph{\scriptsize{Contextual Threat event}} & \multicolumn{2}{c|}{\emph{\scriptsize{Riskier conditions}}} & \multicolumn{2}{c|}{\emph{\scriptsize{Normal conditions}}} & \multicolumn{2}{c|}{\emph{\scriptsize{Riskier conditions}}} & \multicolumn{2}{c|}{\emph{\scriptsize{Normal conditions}}}\tabularnewline
\hline 
\emph{\scriptsize{Defender's Respond and Recovery decision}} & \emph{\scriptsize{Continue drilling}} & \emph{\scriptsize{Stop drilling}} & \emph{\scriptsize{Continue drilling}} & \emph{\scriptsize{Stop drilling}} & \emph{\scriptsize{Continue drilling}} & \emph{\scriptsize{Stop drilling}} & \emph{\scriptsize{Continue drilling}} & \emph{\scriptsize{Stop drilling}}\tabularnewline
\hline 
\hline 
\emph{\scriptsize{Lossing 0 \$ event}} & {\scriptsize{3\%}} & {\scriptsize{0\%}} & {\scriptsize{10\%}} & {\scriptsize{0\%}} & {\scriptsize{92\%}} & {\scriptsize{0\%}} & {\scriptsize{96\%}} & {\scriptsize{0\%}}\tabularnewline
\hline 
\emph{\scriptsize{Lossing 0 - 1 Million \$ event}} & {\scriptsize{12\%}} & {\scriptsize{85\%}} & {\scriptsize{20\%}} & {\scriptsize{90\%}} & {\scriptsize{7\%}} & {\scriptsize{97\%}} & {\scriptsize{4\%}} & {\scriptsize{99\%}}\tabularnewline
\hline 
\emph{\scriptsize{Lossing 1 - 5 Million \$ event}} & {\scriptsize{85\%}} & {\scriptsize{15\%}} & {\scriptsize{70\%}} & {\scriptsize{10\%}} & {\scriptsize{1\%}} & {\scriptsize{3\%}} & {\scriptsize{0\%}} & {\scriptsize{1\%}}\tabularnewline
\hline 
\end{tabular}{\scriptsize \par}

\textbf{\scriptsize{\\Table T4}}{\scriptsize{. Probability table
for UH node.}}{\scriptsize \par}

{\scriptsize{}}%
\begin{tabular}{|>{\centering}p{3.5cm}|>{\centering}p{1.1cm}|>{\centering}p{1.1cm}|>{\centering}p{1.1cm}|>{\centering}p{1.1cm}|>{\centering}p{1.1cm}|>{\centering}p{1.1cm}|>{\centering}p{1.1cm}|>{\centering}p{1.1cm}|}
\hline 
\emph{\scriptsize{Attack event}} & \multicolumn{4}{c|}{\emph{\scriptsize{Attack}}} & \multicolumn{4}{c|}{\emph{\scriptsize{No attack}}}\tabularnewline
\hline 
\emph{\scriptsize{Contextual Threat event}} & \multicolumn{2}{c|}{\emph{\scriptsize{Riskier conditions}}} & \multicolumn{2}{c|}{\emph{\scriptsize{Normal conditions}}} & \multicolumn{2}{c|}{\emph{\scriptsize{Riskier conditions}}} & \multicolumn{2}{c|}{\emph{\scriptsize{Normal conditions}}}\tabularnewline
\hline 
\emph{\scriptsize{Defender's Respond and Recovery decision}} & \emph{\scriptsize{Continue drilling}} & \emph{\scriptsize{Stop drilling}} & \emph{\scriptsize{Continue drilling}} & \emph{\scriptsize{Stop drilling}} & \emph{\scriptsize{Continue drilling}} & \emph{\scriptsize{Stop drilling}} & \emph{\scriptsize{Continue drilling}} & \emph{\scriptsize{Stop drilling}}\tabularnewline
\hline 
\hline 
\emph{\scriptsize{Non casualties event}} & {\scriptsize{96\%}} & {\scriptsize{99.2\%}} & {\scriptsize{99.4\%}} & {\scriptsize{99.96\%}} & {\scriptsize{99.6\%}} & {\scriptsize{99.96\%}} & {\scriptsize{99.9\%}} & {\scriptsize{99.99\%}}\tabularnewline
\hline 
\emph{\scriptsize{Casualties event}} & {\scriptsize{4\%}} & {\scriptsize{0.8\%}} & {\scriptsize{0.6\%}} & {\scriptsize{0.04\%}} & {\scriptsize{0.4\%}} & {\scriptsize{0.04\%}} & {\scriptsize{0.1\%}} & {\scriptsize{0.01\%}}\tabularnewline
\hline 
\end{tabular}{\scriptsize \par}

\textbf{\scriptsize{\\Table T5}}{\scriptsize{. Probability table
for URH node.}}{\scriptsize \par}

{\scriptsize{}}%
\begin{tabular}{|c|c|c|c|c|c|c|}
\hline 
\emph{\scriptsize{Human Consequences event}} & \multicolumn{3}{c|}{\emph{\scriptsize{No casualties}}} & \multicolumn{3}{c|}{\emph{\scriptsize{Casualties}}}\tabularnewline
\hline 
\emph{\scriptsize{Defender's Residual Risk Treatment decision}} & \emph{\scriptsize{Avoid}} & \emph{\scriptsize{Share}} & \emph{\scriptsize{Accept}} & \emph{\scriptsize{Avoid}} & \emph{\scriptsize{Share}} & \emph{\scriptsize{Accept}}\tabularnewline
\hline 
\hline 
\emph{\scriptsize{No casualties event}} & {\scriptsize{99.95\%}} & {\scriptsize{100\%}} & {\scriptsize{100\%}} & {\scriptsize{0\%}} & {\scriptsize{0\%}} & {\scriptsize{0\%}}\tabularnewline
\hline 
\emph{\scriptsize{casualties event}} & {\scriptsize{0.05\%}} & {\scriptsize{0\%}} & {\scriptsize{0\%}} & {\scriptsize{100\%}} & {\scriptsize{100\%}} & {\scriptsize{100\%}}\tabularnewline
\hline 
\end{tabular}{\scriptsize \par}

\textbf{\scriptsize{\\Table T6}}{\scriptsize{. Probability table
for UCA node.}}{\scriptsize \par}

{\scriptsize{}}%
\begin{tabular}{|c|c|c|c|c|}
\hline 
\emph{\scriptsize{Attack event}} & \multicolumn{2}{c|}{{\scriptsize{Attack}}} & \multicolumn{2}{c|}{{\scriptsize{No attack}}}\tabularnewline
\hline 
\emph{\scriptsize{Defender's Forensic System decision}} & \emph{\scriptsize{Forensic}} & \emph{\scriptsize{No forensic}} & \emph{\scriptsize{Forensic}} & \emph{\scriptsize{No forensic}}\tabularnewline
\hline 
\hline 
\emph{\scriptsize{No identification event}} & {\scriptsize{30\%}} & {\scriptsize{90\%}} & {\scriptsize{100\%}} & {\scriptsize{100\%}}\tabularnewline
\hline 
\emph{\scriptsize{Identification event}} & {\scriptsize{70\%}} & {\scriptsize{10\%}} & {\scriptsize{0\%}} & {\scriptsize{0\%}}\tabularnewline
\hline 
\end{tabular}{\scriptsize \par}

\clearpage

\textbf{\scriptsize{\\Table T7}}{\scriptsize{. Probability table
for DC node:}}{\scriptsize \par}

{\scriptsize{}}%
\begin{tabular}{|c|c|c|c|c|}
\hline 
\emph{\scriptsize{Avoiding the risk}} & \multicolumn{4}{c|}{{\scriptsize{10,000,000 \$}}}\tabularnewline
\hline 
\emph{\scriptsize{Sharing the risk}} & \multicolumn{4}{c|}{{\scriptsize{500,000 \$}}}\tabularnewline
\hline 
\multirow{2}{*}{\emph{\scriptsize{Accepting the risk}}} & \emph{\scriptsize{Monetary Consequences event}} & {\scriptsize{0 \$}} & {\scriptsize{0 - 1,000,000 \$}} & {\scriptsize{1,000,000 - 5,000,000 \$}}\tabularnewline
\cline{2-5} 
 & \emph{\scriptsize{Value assigned}} & {\scriptsize{0 \$}} & {\scriptsize{500,000\$}} & {\scriptsize{2,500,000 \$}}\tabularnewline
\hline 
\emph{\scriptsize{Additional protection}} & \multicolumn{4}{c|}{{\scriptsize{20,000 \$}}}\tabularnewline
\hline 
\emph{\scriptsize{Forensic system}} & \multicolumn{4}{c|}{{\scriptsize{10,000 \$}}}\tabularnewline
\hline 
\emph{\scriptsize{Stop drilling}} & \multicolumn{4}{c|}{{\scriptsize{300,000 \$}}}\tabularnewline
\hline 
\end{tabular}{\scriptsize \par}

\textbf{\scriptsize{\\Table T8}}{\scriptsize{. Weight table for DU
node.}}{\scriptsize \par}

{\scriptsize{}}%
\begin{tabular}{|c|c|}
\hline 
\emph{\scriptsize{Importance of the Costs}} & {\scriptsize{5\%}}\tabularnewline
\hline 
\emph{\scriptsize{Importance of the Human Consequences}} & {\scriptsize{95\%}}\tabularnewline
\hline 
\end{tabular}{\scriptsize \par}

\textbf{\scriptsize{\\Table T9}}{\scriptsize{. Value table for ACV
node:}}{\scriptsize \par}

{\scriptsize{}}%
\begin{tabular}{|c|c|c|c|c|}
\hline 
\emph{\scriptsize{Attacker Cost event}} & \multicolumn{2}{c|}{\emph{\scriptsize{Cost}}} & \multicolumn{2}{c|}{\emph{\scriptsize{No cost}}}\tabularnewline
\hline 
\emph{\scriptsize{Counter Attack Consequences event}} & \emph{\scriptsize{No identification}} & \emph{\scriptsize{Identification}} & \emph{\scriptsize{No identification}} & \emph{\scriptsize{Identification}}\tabularnewline
\hline 
\hline 
{\scriptsize{Value}} & {\scriptsize{0.75}} & {\scriptsize{0}} & {\scriptsize{1}} & {\scriptsize{0.25}}\tabularnewline
\hline 
\end{tabular}{\scriptsize \par}

\textbf{\scriptsize{\\Table T10}}{\scriptsize{. Weight table for
AU node.}}{\scriptsize \par}

{\scriptsize{}}%
\begin{tabular}{|c|c|}
\hline 
\emph{\scriptsize{Importance of the costs}} & {\scriptsize{3\%}}\tabularnewline
\hline 
\emph{\scriptsize{Importance of the Monetary Consequences on the Defender}} & {\scriptsize{97\%}}\tabularnewline
\hline 
\end{tabular}{\scriptsize \par}

\textbf{\scriptsize{\\Table T11}}{\scriptsize{. Attacker expected utilities
(in black the highest among the different Attacker's decisions).}}{\scriptsize \par}

{\scriptsize{}}%
\begin{tabular}{|c|c|c|c|>{\centering}m{1.2cm}|>{\centering}m{1.2cm}|>{\centering}m{1.2cm}|>{\centering}m{1.2cm}|}
\hline 
\multirow{2}{*}{\emph{\scriptsize{DP node}}} & \multirow{2}{*}{\emph{\scriptsize{DF node}}} & \multirow{2}{*}{\emph{\scriptsize{DT node}}} & \multirow{2}{*}{\emph{\scriptsize{UC node}}} & \multicolumn{2}{c|}{\emph{\scriptsize{Defender continues drilling}}} & \multicolumn{2}{c|}{\emph{\scriptsize{Defender stops drilling}}}\tabularnewline
\cline{5-8} 
 &  &  &  & \emph{\scriptsize{Perpetrate decision}} & \emph{\scriptsize{Non perpetrate decision}} & \emph{\scriptsize{Perpetrate decision}} & \emph{\scriptsize{Non perpetrate decision}}\tabularnewline
\hline 
\hline 
\multirow{12}{*}{\emph{\scriptsize{Additional protection}}} & \multirow{6}{*}{\emph{\scriptsize{Forensic}}} & \multirow{2}{*}{\emph{\scriptsize{Avoid}}} & \emph{\scriptsize{Riskier conditions}} & \multicolumn{4}{c|}{\textbf{\scriptsize{1}}}\tabularnewline
\cline{4-8} 
 &  &  & \emph{\scriptsize{Normal conditions}} & \multicolumn{4}{c|}{\textbf{\scriptsize{1}}}\tabularnewline
\cline{3-8} 
 &  & \multirow{2}{*}{\emph{\scriptsize{Share}}} & \emph{\scriptsize{Riskier conditions}} & {\scriptsize{0.56074}} & \textbf{\scriptsize{0.56903}} & {\scriptsize{0.61138}} & \textbf{\scriptsize{0.61966}}\tabularnewline
\cline{4-8} 
 &  &  & \emph{\scriptsize{Normal conditions}} & {\scriptsize{0.56074}} & \textbf{\scriptsize{0.56903}} & {\scriptsize{0.61138}} & \textbf{\scriptsize{0.61966}}\tabularnewline
\cline{3-8} 
 &  & \multirow{2}{*}{\emph{\scriptsize{Accept}}} & \emph{\scriptsize{Riskier conditions}} & {\scriptsize{0.36484}} & {\scriptsize{0.35433}} & {\scriptsize{0.61728}} & {\scriptsize{0.62458}}\tabularnewline
\cline{4-8} 
 &  &  & \emph{\scriptsize{Normal conditions}} & {\scriptsize{0.35170}} & {\scriptsize{0.34293}} & {\scriptsize{0.61375}} & {\scriptsize{0.62130}}\tabularnewline
\cline{2-8} 
 & \multirow{6}{*}{\emph{\scriptsize{No forensic}}} & \multirow{2}{*}{\emph{\scriptsize{Avoid}}} & \emph{\scriptsize{Riskier conditions}} & \multicolumn{4}{c|}{\textbf{\scriptsize{1}}}\tabularnewline
\cline{4-8} 
 &  &  & \emph{\scriptsize{Normal conditions}} & \multicolumn{4}{c|}{\textbf{\scriptsize{1}}}\tabularnewline
\cline{3-8} 
 &  & \multirow{2}{*}{\emph{\scriptsize{Share}}} & \emph{\scriptsize{Riskier conditions}} & {\scriptsize{0.55938}} & \textbf{\scriptsize{0.56699}} & {\scriptsize{0.61060}} & \textbf{\scriptsize{0.61821}}\tabularnewline
\cline{4-8} 
 &  &  & \emph{\scriptsize{Normal conditions}} & {\scriptsize{0.55938}} & \textbf{\scriptsize{0.56699}} & {\scriptsize{0.61060}} & \textbf{\scriptsize{0.61821}}\tabularnewline
\cline{3-8} 
 &  & \multirow{2}{*}{\emph{\scriptsize{Accept}}} & \emph{\scriptsize{Riskier conditions}} & {\scriptsize{0.34461}} & {\scriptsize{0.33241}} & {\scriptsize{0.61653}} & {\scriptsize{0.62315}}\tabularnewline
\cline{4-8} 
 &  &  & \emph{\scriptsize{Normal conditions}} & {\scriptsize{0.33055}} & {\scriptsize{0.32013}} & {\scriptsize{0.61299}} & {\scriptsize{0.61986}}\tabularnewline
\hline 
\multirow{12}{*}{\emph{\scriptsize{No additional protection}}} & \multirow{6}{*}{\emph{\scriptsize{Forensic}}} & \multirow{2}{*}{\emph{\scriptsize{Avoid}}} & \emph{\scriptsize{Riskier conditions}} & \multicolumn{4}{c|}{{\scriptsize{1}}}\tabularnewline
\cline{4-8} 
 &  &  & \emph{\scriptsize{Normal conditions}} & \multicolumn{4}{c|}{{\scriptsize{1}}}\tabularnewline
\cline{3-8} 
 &  & \multirow{2}{*}{\emph{\scriptsize{Share}}} & \emph{\scriptsize{Riskier conditions}} & {\scriptsize{0.55116}} & \textbf{\scriptsize{0.56496}} & {\scriptsize{0.60295}} & \textbf{\scriptsize{0.61675}}\tabularnewline
\cline{4-8} 
 &  &  & \emph{\scriptsize{Normal conditions}} & {\scriptsize{0.55116}} & \textbf{\scriptsize{0.56496}} & {\scriptsize{0.60295}} & \textbf{\scriptsize{0.61675}}\tabularnewline
\cline{3-8} 
 &  & \multirow{2}{*}{\emph{\scriptsize{Accept}}} & \emph{\scriptsize{Riskier conditions}} & {\scriptsize{0.45634}} & {\scriptsize{0.29898}} & {\scriptsize{0.61588}} & {\scriptsize{0.62173}}\tabularnewline
\cline{4-8} 
 &  &  & \emph{\scriptsize{Normal conditions}} & {\scriptsize{0.42794}} & {\scriptsize{0.28532}} & {\scriptsize{0.61058}} & {\scriptsize{0.61841}}\tabularnewline
\cline{2-8} 
 & \multirow{6}{*}{\emph{\scriptsize{No Forensic}}} & \multirow{2}{*}{\emph{\scriptsize{Avoid}}} & \emph{\scriptsize{Riskier conditions}} & \multicolumn{4}{c|}{{\scriptsize{1}}}\tabularnewline
\cline{4-8} 
 &  &  & \emph{\scriptsize{Normal conditions}} & \multicolumn{4}{c|}{{\scriptsize{1}}}\tabularnewline
\cline{3-8} 
 &  & \multirow{2}{*}{\emph{\scriptsize{Share}}} & \emph{\scriptsize{Riskier conditions}} & {\scriptsize{0.55442}} & \textbf{\scriptsize{0.56282}} & {\scriptsize{0.60690}} & \textbf{\scriptsize{0.61530}}\tabularnewline
\cline{4-8} 
 &  &  & \emph{\scriptsize{Normal conditions}} & {\scriptsize{0.55442}} & \textbf{\scriptsize{0.56282}} & {\scriptsize{0.60690}} & \textbf{\scriptsize{0.61530}}\tabularnewline
\cline{3-8} 
 &  & \multirow{2}{*}{\emph{\scriptsize{Accept}}} & \emph{\scriptsize{Riskier conditions}} & {\scriptsize{0.32392}} & {\scriptsize{0.07465}} & {\scriptsize{0.61990}} & {\scriptsize{0.62030}}\tabularnewline
\cline{4-8} 
 &  &  & \emph{\scriptsize{Normal conditions}} & {\scriptsize{0.28286}} & {\scriptsize{0.05131}} & {\scriptsize{0.61456}} & {\scriptsize{0.61696}}\tabularnewline
\hline 
\end{tabular}{\scriptsize \par}

\clearpage

\textbf{\scriptsize{\\Table T12}}{\scriptsize{. Defender expected
utilities (in black the highest among the different Defender's decisions).}}{\scriptsize \par}

{\scriptsize{}}%
\begin{tabular}{|c|c|c|c|>{\centering}p{1.4cm}|>{\centering}p{1.4cm}|>{\centering}p{1.4cm}|>{\centering}p{1.4cm}|}
\hline 
\multirow{3}{*}{\emph{\scriptsize{DP node}}} & \multirow{3}{*}{\emph{\scriptsize{DF node}}} & \multirow{3}{*}{\emph{\scriptsize{DT node}}} & \multirow{3}{*}{\emph{\scriptsize{DR node}}} & \multicolumn{4}{c|}{\emph{\scriptsize{Possible events}}}\tabularnewline
\cline{5-8} 
 &  &  &  & \multicolumn{2}{c|}{\emph{\scriptsize{Riskier conditions}}} & \multicolumn{2}{c|}{\emph{\scriptsize{Normal conditions}}}\tabularnewline
\cline{5-8} 
 &  &  &  & \emph{\scriptsize{Attack event}} & \emph{\scriptsize{Non attack event}} & \emph{\scriptsize{Attack event}} & \emph{\scriptsize{Non attack event}}\tabularnewline
\hline 
\multirow{12}{*}{\emph{\scriptsize{Additional protection}}} & \multirow{6}{*}{\emph{\scriptsize{Forensic}}} & \multirow{2}{*}{\emph{\scriptsize{Avoid}}} & \emph{\scriptsize{Continue drilling}} & {\scriptsize{0.91154}} & {\scriptsize{0.94573}} & {\scriptsize{0.94383}} & {\scriptsize{0.94858}}\tabularnewline
\cline{4-8} 
 &  &  & \emph{\scriptsize{Stop drilling}} & {\scriptsize{0.94193}} & {\scriptsize{0.94915}} & {\scriptsize{0.94915}} & {\scriptsize{0.94943}}\tabularnewline
\cline{3-8} 
 &  & \multirow{2}{*}{\emph{\scriptsize{Share}}} & \emph{\scriptsize{Continue drilling}} & {\scriptsize{0.95935}} & {\scriptsize{0.99355}} & {\scriptsize{0.99165}} & {\scriptsize{0.99640}}\tabularnewline
\cline{4-8} 
 &  &  & \emph{\scriptsize{Stop drilling}} & {\scriptsize{0.98825}} & {\scriptsize{0.99547}} & {\scriptsize{0.99547}} & {\scriptsize{0.99576}}\tabularnewline
\cline{3-8} 
 &  & \multirow{2}{*}{\emph{\scriptsize{Accept}}} & \emph{\scriptsize{Continue drilling}} & {\scriptsize{0.95092}} & {\scriptsize{0.99575}} & {\scriptsize{0.98490}} & {\scriptsize{0.99880}}\tabularnewline
\cline{4-8} 
 &  &  & \emph{\scriptsize{Stop drilling}} & {\scriptsize{0.98675}} & {\scriptsize{0.99517}} & {\scriptsize{0.99447}} & {\scriptsize{0.99566}}\tabularnewline
\cline{2-8} 
 & \multirow{6}{*}{\emph{\scriptsize{No forensic}}} & \multirow{2}{*}{\emph{\scriptsize{Avoid}}} & \emph{\scriptsize{Continue drilling}} & {\scriptsize{0.91154}} & {\scriptsize{0.94573}} & {\scriptsize{0.94383}} & {\scriptsize{0.94858}}\tabularnewline
\cline{4-8} 
 &  &  & \emph{\scriptsize{Stop drilling}} & {\scriptsize{0.94193}} & {\scriptsize{0.94915}} & {\scriptsize{0.94915}} & {\scriptsize{0.94943}}\tabularnewline
\cline{3-8} 
 &  & \multirow{2}{*}{\emph{\scriptsize{Share}}} & \emph{\scriptsize{Continue drilling}} & {\scriptsize{0.95940}} & {\scriptsize{0.99360}} & {\scriptsize{0.99170}} & {\scriptsize{0.99645}}\tabularnewline
\cline{4-8} 
 &  &  & \emph{\scriptsize{Stop drilling}} & {\scriptsize{0.98830}} & {\scriptsize{0.99552}} & {\scriptsize{0.99552}} & {\scriptsize{0.99581}}\tabularnewline
\cline{3-8} 
 &  & \multirow{2}{*}{\emph{\scriptsize{Accept}}} & \emph{\scriptsize{Continue drilling}} & {\scriptsize{0.95097}} & {\scriptsize{0.99580}} & {\scriptsize{0.98495}} & {\scriptsize{0.99885}}\tabularnewline
\cline{4-8} 
 &  &  & \emph{\scriptsize{Stop drilling}} & {\scriptsize{0.98680}} & {\scriptsize{0.99522}} & {\scriptsize{0.99452}} & {\scriptsize{0.99571}}\tabularnewline
\hline 
\multirow{12}{*}{\emph{\scriptsize{No additional protection}}} & \multirow{6}{*}{\emph{\scriptsize{Forensic}}} & \multirow{2}{*}{\emph{\scriptsize{Avoid}}} & \emph{\scriptsize{Continue drilling}} & {\scriptsize{0.91154}} & {\scriptsize{0.94573}} & {\scriptsize{0.94383}} & {\scriptsize{0.94858}}\tabularnewline
\cline{4-8} 
 &  &  & \emph{\scriptsize{Stop drilling}} & {\scriptsize{0.94193}} & {\scriptsize{0.94915}} & {\scriptsize{0.94915}} & {\scriptsize{0.94943}}\tabularnewline
\cline{3-8} 
 &  & \multirow{2}{*}{\emph{\scriptsize{Share}}} & \emph{\scriptsize{Continue drilling}} & {\scriptsize{0.95945}} & {\scriptsize{0.99365}} & {\scriptsize{0.99175}} & {\scriptsize{0.99650}}\tabularnewline
\cline{4-8} 
 &  &  & \emph{\scriptsize{Stop drilling}} & {\scriptsize{0.98835}} & {\scriptsize{0.99557}} & {\scriptsize{0.99557}} & {\scriptsize{0.99586}}\tabularnewline
\cline{3-8} 
 &  & \multirow{2}{*}{\emph{\scriptsize{Accept}}} & \emph{\scriptsize{Continue drilling}} & {\scriptsize{0.95102}} & {\scriptsize{0.99585}} & {\scriptsize{0.98500}} & {\scriptsize{0.99890}}\tabularnewline
\cline{4-8} 
 &  &  & \emph{\scriptsize{Stop drilling}} & {\scriptsize{0.98685}} & {\scriptsize{0.99527}} & {\scriptsize{0.99457}} & {\scriptsize{0.99576}}\tabularnewline
\cline{2-8} 
 & \multirow{6}{*}{\emph{\scriptsize{No Forensic}}} & \multirow{2}{*}{\emph{\scriptsize{Avoid}}} & \emph{\scriptsize{Continue drilling}} & {\scriptsize{0.91154}} & {\scriptsize{0.94573}} & {\scriptsize{0.94383}} & {\scriptsize{0.94858}}\tabularnewline
\cline{4-8} 
 &  &  & \emph{\scriptsize{Stop drilling}} & {\scriptsize{0.94193}} & {\scriptsize{0.94915}} & {\scriptsize{0.94915}} & {\scriptsize{0.94943}}\tabularnewline
\cline{3-8} 
 &  & \multirow{2}{*}{\emph{\scriptsize{Share}}} & \emph{\scriptsize{Continue drilling}} & {\scriptsize{0.95950}} & {\scriptsize{0.99370}} & {\scriptsize{0.99180}} & {\scriptsize{0.99655}}\tabularnewline
\cline{4-8} 
 &  &  & \emph{\scriptsize{Stop drilling}} & \textbf{\scriptsize{0.98840}} & {\scriptsize{0.99562}} & \textbf{\scriptsize{0.99562}} & {\scriptsize{0.99591}}\tabularnewline
\cline{3-8} 
 &  & \multirow{2}{*}{\emph{\scriptsize{Accept}}} & \emph{\scriptsize{Continue drilling}} & {\scriptsize{0.95107}} & \textbf{\scriptsize{0.99590}} & {\scriptsize{0.98505}} & \textbf{\scriptsize{0.99895}}\tabularnewline
\cline{4-8} 
 &  &  & \emph{\scriptsize{Stop drilling}} & {\scriptsize{0.98690}} & {\scriptsize{0.99532}} & {\scriptsize{0.99462}} & {\scriptsize{0.99581}}\tabularnewline
\hline 
\end{tabular}{\scriptsize \par}

\end{center}
\end{document}